\documentclass[aps,twocolumn,showpacs,preprintnumbers,prb]{revtex4-1}
\usepackage{dcolumn}
\usepackage{graphicx}
\usepackage{amsmath, amssymb}
\usepackage{bm}
\usepackage{mathrsfs}

\begin{document}

\title{Field theory analysis of $S=1$ antiferromagnetic
bond-alternating chains \\ in the dimer phase}

\author{Junya Tamaki}

\affiliation{ Department of Physics, Tokyo Institute
of Technology, Oh-okayama, Meguro-ku, Tokyo 152-8551, Japan}

\author{Masaki Oshikawa}\thanks{Corresponding author}

\affiliation{Institute for
Solid State Physics, University of Tokyo, Kashiwa 277-8581, Japan}

\begin{abstract}
Dynamics of
$S=1$ antiferromagnetic bond-alternating chains in the 
dimer phase, in the vicinity of the critical point with
the Haldane phase, is studied by a field theoretical method.
This model is considered to represent the
compound Ni(C$_9$H$_{24}$N$_4$)(NO$_2$)ClO$_4$
(abbreviated as NTENP).
We derive a sine-Gordon (SG) field theory as a
low-energy effective model of this system, starting from a
Tomonaga-Luttinger liquid at the critical point.  Using the exact
solution of the SG theory, we give a field theoretical picture of the
low-energy excitation spectrum of NTENP.  Results derived from our
picture are in a good agreement with results of inelastic neutron
scattering experiments on NTENP and numerical calculation of the
dynamical structure factor.
Furthermore, on the basis of the obtained theoretical picture, we predict
that the sharp peaks correspond to a single elementary excitation are
absent in the Raman scattering spectrum of NTENP in contrast to the
inelastic neutron scattering spectrum.
\end{abstract}
\date{\today}

\maketitle

\section{Introduction} 
One dimensional antiferromagnetic Heisenberg chains show various
physical properties under strong quantum fluctuations.  They have
offered many fascinating phenomena as a stage of both theoretical and
experimental studies for many years. 
While an uniform antiferromagnetic Heisenberg chain is gapless for
half-integer spins, it has an exotic ground state with a finite gap for
integer spins.~\cite{haldane} This gapped phase is called Haldane phase.
The antiferromagnetic Heisenberg chain is not exactly solvable except
for $S=1/2$.  However, the Haldane phase could be understood
in terms of the solvable
Affleck-Kennedy-Lieb-Tasaki model~\cite{affleck,aklt}.
Its exact groundstate can be constructed in terms of
valence bonds (singlet pairs of constituent $S=1/2$'s).

Bond-alternating antiferromagnetic Heisenberg chain defined by
\begin{equation}
\mathcal{H} = J \sum_j ( \bm{S} _{2j-1} \cdot \bm{S} _{2j} + \alpha \,
\bm{S}_{2j} \cdot \bm{S}_{2j+1} ), \label{eq:BondAlt}
\end{equation}
where $\alpha$ represents the bond alternation ratio, is an interesting
generalization of the Haldane gap problem.  In this model, for spin
quantum number $S$, the system shows $2S$ successive quantum phase
transitions in $0 < \alpha < \infty$~\cite{afflecknucl,affleckhaldane}.
This may be understood, in the AKLT picture, as follows.  In the limit of
$\alpha=0$, all the valence bonds are placed on the ``odd'' links
between sites $2j-1,2j$.
In the opposite limit $\alpha=\infty$, all the
valence bonds are placed on the ``even'' links between sites $2j, 2j+1$.
They
correspond to completely dimerized states.  By changing the number of
valence bonds on odd and even links, we can construct $2S+1$ AKLT-type
states with varying degree of dimerization.  Each of these AKLT-type
states represents the $2S+1$ gapped phases, separated by the $2S$
quantum phase transitions.  Each of the transitions may be regarded as
a rearrangement of valence bonds, in which one valence bond is
transferred between neighboring links.

In the case of an integer spin $S$, $2S+1$ gapped phases can be
classified into two categories: one with an even number of valence bonds on
every link, and the other with an odd number of valence bonds.  The
latter is a (symmetry-protected) topological phase; in the presence of
global $Z_2 \times Z_2$ symmetry of $\pi$-rotation about $x,y$ and $z$
axes, it is characterized by spontaneous breaking of a {\em hidden} $Z_2
\times Z_2$ symmetry~\cite{kennedytasaki,Oshikawa1992}, which can be
detected by the string order parameter.~\cite{nijsrommelse} More
generally, the topological phase is also protected by either the
time-reversal symmetry or the lattice inversion symmetry about a link,
and is characterized by exact two-fold degeneracy of entire entanglement
spectrum.~\cite{Pollmann2009,Pollmann-PRB2010}

In this paper, we focus on the simple case of $S=1$.  Here, the $2S+1=3$
gapped phases consist of two dimer phases (at large and small values of
$\alpha$) and the $S=1$ Haldane phase in a range of $\alpha$ including
$\alpha=1$.  As mentioned above, being an exotic ``topological phase'',
the Haldane state has been studied vigorously.  In contrast, the dimer
phase may be regarded as a trivial phase as far as the ground-state
properties are concerned, and has been relatively less studied.  In
fact, the system is exactly solvable in the dimer limit $\alpha =0$,
where the system is reduced to a two-body problem.  However, dynamical
properties, which are related to excited states, are not necessarily
trivial even in the dimer phase.  While quantum phases are usually
classified with respect to order parameters in the ground state, they
are not sufficient for elucidation of the excited states.  Thus the
dimer phase, although without any nontrivial order, has a possibility of
exhibiting a rich structure in the excitations.  In addition, it would
be also interesting to compare dynamical properties between the dimer
phase and the Haldane phase.  Therefore, in this paper, we attempt to
study the dynamical properties in the $S=1$ dimer phase, especially in
the neighborhood of the quantum critical point.


Theoretical study of Haldane gap has been, without doubt, stimulated by
discovery of several model materials, such as
Ni(C$_2$H$_{8}$N$_8$)$_2$(NO$_2$)ClO$_4$ (abbreviated to NENP) and
Ni(C$_5$H$_{14}$N$_2$)$_2$N$_3$(PF$_6$) (abbreviated to NDMAP).
Fortunately, materials representing the $S=1$ bond-alternating
antiferromagnetic chain~\eqref{eq:BondAlt}, such as
Ni(C$_9$H$_{24}$N$_4$)(NO$_2$)ClO$_4$ (abbreviated to NTENP)~\cite{chem}
and [{Ni(333-tet)($\mu$-N$_3$)}$_n$](ClO$_4$)$_n$ were also discovered,
and the bond-alternation ratio $\alpha$ was identified in these
materials.  In the latter system, quite remarkably, $\alpha$ turns out
to be exactly (within experimental accuracy) at the critical point
separating the Haldane and dimer phases.~\cite{gapless}
In NTENP, on the other hand, $\alpha$ is slightly smaller than the critical
value.~\cite{ntenp} That is, the system belongs to the dimer
phase, but in vicinity of the quantum critical point.  Thus
NTENP offers an ideal playground to study nontrivial dynamics in the
dimer phase.

Dynamical properties of NTENP are studied experimentally by inelastic
neutron scattering (INS),~\cite{zheludev,regnault,hagiwara} and are
numerically studied by using a continued-fraction method based on the
Lanczos algorithm \cite{suzukisuga}.  However, in order to better
understand the dynamics in NTENP and also in related materials in a
unified way, it would be desirable to develop a coherent theoretical
picture.  In this paper, we present a field theoretical approach to the
problem, which we hope would serve the purpose.

At a quantum critical point, physical quantities are often constrained
by the conformal invariance, and the system is described by a conformal
field theory (CFT).  Near the critical point, the system could be
understood by a CFT with a small (but relevant) perturbation.  In this
paper, we follow this strategy, making use of the feature of NTENP that
the bond-alternating ratio is in the vicinity of the critical point.
Based on the obtained effective theory, we elucidate the dynamical
properties of NTENP and discuss preceding experimental and numerical
results. Furthermore, we make predictions on Raman Scattering experiment
which has not yet been carried out.

\section{Model and Method}

In fact, NTENP has a sizable spin anisotropy; it is better described,
instead of eq.~\eqref{eq:BondAlt}, by the following
Hamiltonian~\cite{ntenp}
\begin{equation}
\mathcal{H} = J \sum_j ( \bm{S} _{2j-1} \cdot \bm{S} _{2j} + \alpha \,
\bm{S}_{2j} \cdot \bm{S}_{2j+1} ) + D \sum_j(S_j^z)^2 ,
\label{eq:hamiltonian}
\end{equation}
where $J > 0$, $\alpha$ and $D$ represent bond alternation and uniaxial
anisotropy, respectively.  For NTENP, these parameters are $D \sim
0.25J$ and $\alpha \sim 0.45$.
The spin anisotropy axis (which is
determined by the chain axis) is taken to be $z$-direction.
In this paper, we ignore the in-plane anisotropy
$E \sum_j (S^x_j)^2 - (S^y_j)^2$ and the exchange anisotropy,
which are expected to be smaller than $D$ in NTENP.
The lattice
constant between neighboring two spins is set to unity.

The critical point $\alpha_c$, where energy gap is closed, is about 0.6
at $D=0$.  It is a function of $D$ which is numerically evaluated in
Ref.~\onlinecite{chenhida}. According to them, $\alpha_c$ is almost constant
for $0 < D \lesssim 0.3 J$.  In NTENP, $D \sim 0.25 J $ and thus we can
assume that $\alpha_c \sim 0.6$.  The actual bond-alternating ratio of
NTENP is $\alpha \sim 0.45 < \alpha_c$, and the system is certainly in
the dimer phase.  However, it can be still regarded as
a vicinity of the quantum critical point.

At the critical point $\alpha = \alpha_c$, the system is mapped to a TL
liquid,~\cite{tl} in the low-energy limit.~\cite{ntenp} It is nothing
but the theory of free boson field $\phi$ with the Lagrangian density,
\begin{equation}
\mathcal{L}=\frac{1}{2}(\partial_{\mu}\phi)^2 .
\label{eq:tl}\end{equation} Let $R$ be the compactification radius of
boson field,
\begin{equation}
\phi(x,t)+ 2\pi R \sim \phi(x,t) .
\label{eq:compactification}
\end{equation}
The compactification radius
determines critical exponents of the TL liquid.  This may be understood
as a consequence of ``quantization'' of vertex operators required by the
compactification. For example, the vertex operators $\cos{(\gamma
\phi)}$ must have the coefficient $\gamma = n/R$ with an integer
$n$, in order to be single-valued under the
compactification~\eqref{eq:compactification}.
The scaling dimension of the vertex operator
$\cos{(n \phi /R)}$ is known\cite{tl} as
\begin{equation}
 x_n = \frac{n^2}{4 \pi R^2}.
\end{equation}
The operator is relevant (irrelevant) in the Renormalization-Group
(RG) sense, if the scaling dimension is smaller (greater) than $2$.
Thus, among the vertex operators of the above form,
$n=1$ is the most relevant.
As we will discuss later, in the present application
$R$ is somewhat smaller than the SU(2) symmetric value
$1/\sqrt{2\pi}$, implying $n=1$ is the only relevant operator in
the family $\cos{(n \phi /R)}$.

When $\alpha \neq \alpha_c$, the system acquires a gap and it is no
longer described by the TL liquid (\ref{eq:tl}).  However, if $|\alpha -
\alpha_c|$ is small, as it is the case in NTENP, the low-energy
effective theory would be given by the TL liquid with a perturbation.
We postulate that the leading perturbation is the most relevant operator
permitted under the symmetry of the Hamiltonian (\ref{eq:hamiltonian}) :
$\cos{(\phi/R)}$.  Hence the low-energy effective theory for small
$|\alpha - \alpha_c|$ should be following sine-Gordon (SG) field theory,
\begin{equation}
\mathcal{L}=\frac{1}{2}(\partial_{\mu}\phi)^2 +
C \cos{\big( \frac{\phi}{R} \big)} ,
\label{eq:sine-Gordon}
\end{equation}
where $C$ is the constant that is
proportional to $\alpha - \alpha_c$ and $R$ is the compactification
radius of boson field at the critical point.  When $\alpha = \alpha_c$,
$C$ equals zero and Eq.~(\ref{eq:sine-Gordon}) is reduced to
Eq.~(\ref{eq:tl}).  It should be noted that we use $R$ at the critical
point $\alpha_c$ in this construction (\ref{eq:sine-Gordon}) for $\alpha
\neq \alpha_c$.  The compactification radius $R$ is well defined only at
the critical point and the precise determination of $R$ is subtle for
$\alpha \neq \alpha_c$.  Nevertheless, $R$ obtained at $\alpha =
\alpha_c$ should be a reasonable approximation as long as $|\alpha -
\alpha_c|$ is small.  We also note that $R$ at the critical point
$\alpha_c$ is a function of $D$.

The SG field theory is integrable and its exact solution is
known.~\cite{bergknoff,korepin} In this study, we analyze the low-energy
spectrum of NTENP using the exact solution of the SG field theory, where
the original spin Hamiltonian is mapped.  We thus refer to this exact
solution in the next section.

\section{Exact Solution of the Sine-Gordon Field Theory}

The SG field theory has the Lorentz invariance, hence the energy of the
elementary excitations obey the relativistic dispersion relation
\begin{equation}
\epsilon=\sqrt{k^2 v_s^2+m^2} . 
\label{eq:relativistic_dispersion}
\end{equation}
Here, $v_s$ is the spin-wave velocity which plays the
role of the speed of light, $k$ is the momentum, and
$m$ is the mass of the excitation (in the unit of energy),
which represents the lowest creation
energy of each elementary excitations.

The elementary excitations of the SG field theory (\ref{eq:sine-Gordon})
consist of a soliton, an antisoliton and several breathers. The soliton
and the antisoliton have the same mass $M_S$.
Breathers are bound states of a soliton and an antisoliton.
The exact solution\cite{bergknoff,korepin} of the
SG field theory implies the number of
different kinds of breathers as $[1/\xi]$, where
\begin{equation}
\frac{1}{\xi} = 8 \pi R^2 - 1,  \label{eq:xi}
\end{equation}
and $[x]$ denotes the integer part of $x$.
Remarkably, the exact mass ratio was also derived as
\begin{equation}
\frac{M_n}{M_S} = 2\sin{\big( \frac{n\pi\xi}{2} \big)}
\hspace{4mm} (n<1/\xi) ,
\label{eq:massratio}
\end{equation}
where $M_n$ and $M_S$ are respectively the $n$-th breather mass and the
soliton mass.
Here we note that the breather mass $M_n$ does not exceed
twice the soliton mass $2M_S$.
Physically, this can be interpreted as follows.
If the mass of a breather were larger than $2M_S$, the breather
would decay into free soliton and antisoliton, and would not
constitute a stable elementary excitation.
It should be emphasized
that the mass ratio (\ref{eq:massratio}) and the number of different
breathers depend only on the boson field compactification radius $R$,
which is a function of $D$.

\begin{figure}[tb]
\centering \includegraphics[width=3in,clip]{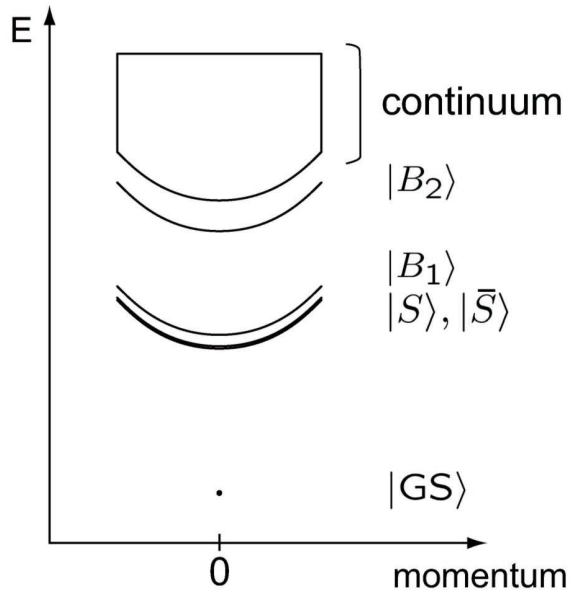} \caption{Schematic
view of the excitation spectrum of the SG field
theory. $|\textrm{GS}\rangle$, $|S\rangle$, $|\bar{S}\rangle$,
$|B_1\rangle$ and $|B_2\rangle$ represent the ground state, the soliton
state, the antisoliton state, the first breather state and the second
breather state, respectively.  The soliton level and the antisoliton
level is degenerated.  The number of different breathers is $[1/\xi]$,
where $[x]$ denotes the integer part of $x$ and $1/\xi = \frac{2}{\pi
R^2} - 1$.  In this schematic view, the case that the theory has two
breathers is represented. 
Owing to the Lorentz invariance of the theory, the excitation
energies obey the relativistic dispersion
relation~\protect\eqref{eq:relativistic_dispersion}.
}
\label{fig:sine-Gordon}
\end{figure}

The above picture of elementary excitations implies
the spectrum of the excited states
schematically described in Fig.~\ref{fig:sine-Gordon}.
Each elementary excitation obeys the relativistic dispersion
(\ref{eq:relativistic_dispersion}) with
a species-dependent mass $m$ and the common
spin-wave velocity $v_s$.
Above the energy $2M_S$, there is a continuum of
excited states, corresponding to scattering states
of two or more elementary excitations.

The elementary excitations of NTENP is understood as the corresponding
elementary excitations of the SG field theory.  The boson field
compactification radius $R$ of NTENP determines the mass ratio between a
soliton and breathers and the number of different breathers correspond
to NTENP.

In passing, we note that,
for an isotropic system, the SU(2) symmetry requires $R=1/\sqrt{2\pi}$.
The exact solution of the SG theory
implies that there are two breathers, and the mass of
the 1st breather is degenerate with soliton/antisoliton,
forming a SU(2) triplet.
Furthermore, the mass of the second breather is $\sqrt{3}$
times that of the triplet.
However, this does not quite agree with
the numerical results.
This is due to the marginal perturbation to the SG theory,
which generally exists in the SU(2) symmetric system.
Only with an appropriate strength of next-nearest neighbor
interaction, the marginal operator can be eliminated
and the above prediction
of the SG theory is realized~\cite{Bouzerar98,Takayoshi-SB}.
In this paper, however, we consider a system with a substantial
anisotropy, where the marginal operator is absent.
Thus the simple SG theory~\eqref{eq:sine-Gordon}
is expected to be a good low-energy effective theory of the system.

\section{Structure of elementary excitations}
\label{sec:result}

\subsection{Mass ratio among the elementary excitations}
\label{subsec:massratio}

We are now in a position to apply the relation
(\ref{eq:massratio}) to the system of NTENP.  Here we note again that
the critical point $\alpha_c$ and the compactification radius $R$ at
this point are a function of the uniaxial anisotropy $D$. At $D=0$, $R$
is determined as $2\pi R^2 =1$ by the $SU(2)$ symmetry of the system.
The variation of $R$ at several values of $D$ is already studied
numerically by using CFT and the level spectroscopy method, by Chen et
al.~\cite{chenhida} The parameter $K$, $\delta$ and $D$ in
Ref.~\onlinecite{chenhida} correspond to $1/(2\pi R^2)$, $(1-\alpha)/(1+\alpha)$
and $2D/(1+\alpha)$ respectively, in the notation of the present paper.
Thus, for NTENP with $D \sim 0.25J$, we find
$1/(2\pi R^2) \sim 1.7$ that implies $1/\xi\sim1.2$.
It follows that this effective SG field theory for NTENP
has the first breather only.
The breather/soliton mass ratio is
\begin{equation}
 b = \frac{M_1}{M_S} = 1.93.
\label{eq.massratio.NTENP}
\end{equation}
Accordingly, the energy spectrum
of NTENP based on the SG field theory picture appears to be the
following: the lowest excited state is a degenerate doublet corresponds
to the soliton and the antisoliton, the next lowest excited state is a
single mode corresponds to the breather, and an excitation continuum
starts from the energy twice the soliton mass.  This spectrum is
schematically described in Fig.~\ref{fig:NTENPspectrum}.

\subsection{Dynamical structure factor}

\label{subsec:dsf}
The dynamical structure factor (DSF) is defined as
the Fourier transform of the correlation function.  It is an important
quantity, as the differential cross section in an INS is proportional to
the DSF.
The $\mu$ ($=x,y,z$) component of the DSF at zero temperature is
\begin{align}
\mathcal{S}^{\mu\mu}(q,\omega) &= \int_{-\infty }^{\infty } dt
\sum_{j,k} \, \langle \psi_0 | S^\mu_j(t)S^\mu_k(0) | \psi_0 \rangle \,
e^{i\omega t -iq(j-k)} \nonumber \\[-3pt] &=\sum_n \delta(\omega - (E_n
- E_0)) \, |\langle \psi_n | S^\mu_q |\psi_0 \rangle |^2 ,
\label{eq:dsf}\end{align} where $|\psi_0 \rangle$ and $|\psi_n \rangle$
are the ground state and the excited state of Hamiltonian
(\ref{eq:hamiltonian}) with the eigenvalue $E_0$ and $E_n$,
respectively, and $S^\mu_q = \frac{1}{\sqrt{2N}} \sum_{j=1}^{2N} S^\mu_j
e^{-iqj}$, where $2N$ is the total number of spins.  Here it should be
noted that, in the systems with a bond alternation such as NTENP, the
width of the Brillouin zone is reduced to half of that of the uniform
chain.  Namely, the conserved crystal momentum can be defined as
\begin{equation}
 \tilde{q} = q \mod{\pi} ,
\end{equation}
which is defined on the reduced Brillouin
zone
\begin{equation}
-\pi/2 \leq \tilde{q} < \pi/2
\label{eq.reducedBZ}
\end{equation}
In particular, $q=0$ and $q=\pi$ are
both identified with the center of the reduced Brillouin zone
$\tilde{q}=0$.
However, it should be noted that, the DSF is {\em not}
equivalent for $q$ and $q+\pi$. This could be easily understood by
considering the limit of zero dimerization, where $q$ and $q+\pi$ are
certainly distinguishable. The operator $S^{\mu}_q$ and
$S^{\mu}_{q+\pi}$ are different operators and thus give different DSFs.
In the presence of dimerization, they are governed by the same
selection rule. The matrix elements, however, are generally different.

The summation in Eq. (\ref{eq:dsf}) is taken over all intermediate
states, but many of the matrix elements vanish due to selection rules.
While excited states $|\psi_n \rangle$ contain one or more elementary
excitations, only the states which have non-vanishing overlap with
$S^\mu_q |\psi_0 \rangle $ contribute to the DSF.  Thus, in order to
evaluate each component of the DSF, it is important to know what kind of
elementary excitations are created by applying $S^\mu_q$ on the ground
state $|\psi_0 \rangle$.  For example, if $S^\mu_q|\psi_0 \rangle $
contains a single soliton state, there must be a contribution
proportional to $\delta(\omega - \sqrt{\tilde{q}^2 + M_S^2})$, to
$\mathcal{S}^{\mu\mu}(q,\omega)$.

\subsection{Dimer limit}
\label{sec.dimer}

In order to clarify which elementary excitation is created by
application of each spin operator,
we need to know the correspondence between the operators in
the SG field theory and those in the original spin system.  In
principle, this would follow from a microscopic derivation of the
effective SG field theory from the spin model~(\ref{eq:hamiltonian}).
However, the microscopic derivation is rather complicated and the
correspondence cannot be easily established.  A possible derivation
starts from the bosonization of two $S=1/2$ chains and then introduces a
strong ferromagnetic coupling between them, to form a $S=1$ chain
effectively\cite{Schulz86}.
Instead of pursuing this direction, we shall develop a simple
ansatz for the correspondence, based on the dimer
limit $\alpha=0$.

In the dimer limit, the problem is reduced to
a two-spin problem, and thus can be exactly solved.
Two $S=1$'s coupled with the Heisenberg antiferromagnetic exchange,
$\mathcal{H} = J \bm{S}_1 \cdot \bm{S}_2$ has
the singlet groundstate with energy $-2J$, triplet excited states
with energy $-J$, quintet excited states with energy $+J$.
Now let us introduce the uniaxial anisotropy $D$ to
the dimer, and consider the Hamiltonian
\begin{equation}
\mathcal{H} = J \bm{S} _1 \cdot \bm{S} _2 +
D \big( (S^z_1)^2 + (S^z_2)^2 \big) .
\end{equation}
The singlet groundstate is perturbed by the anisotropy,
and given by
\begin{equation}
|s\rangle=C_s
\left(
|\uparrow \downarrow \rangle+|\downarrow \uparrow
\rangle+ \frac{4}{A-\sqrt{A^2+8}}|00\rangle
\right),
\end{equation}
with the groundstate energy 
\begin{equation}
\frac{A-\sqrt{A^2+8}}{2} J,
\end{equation}
where $A=2D-1$ and $C_s$ is the normalization
constant. 

The uniaxial anisotropy splits the triplet and quintet
excited states.
Each state can be labelled by
the total magnetization $S^z = S^z_1 + S^z_2$,
which is still a good quantum number.
Since the uniaxial anisotropy does not break
the $\pi$-rotation symmetry under rotation about
$x$-axis.
Thus, the triplet states are split into
a singlet with $S^z=0$, which is denoted by $|t^0\rangle$,
and a doublet $|t^\pm \rangle$ with $S^z=\pm 1$.
The doublet states are given by
\begin{align}
|t^+ \rangle &= \frac{1}{\sqrt{2}}(|\uparrow 0\rangle-|0\uparrow
\rangle), \\
|t^- \rangle &= \frac{1}{\sqrt{2}}(|\downarrow
0\rangle-|0\downarrow \rangle),
\end{align}
with the energy $D - J$,
while the singlet is given by
\begin{equation}
|t^0\rangle=\frac{1}{\sqrt{2}}(|\uparrow \downarrow \rangle-|\downarrow
\uparrow \rangle),
\end{equation}
with the energy $2D - J$.
For $D>0$, which is the case for NTENP, the doublet states $|t^\pm\rangle$
are lower in energy than the singlet $|t^0\rangle$.

In a similar way, the quintet states are split into a singlet with
$S^z=0$ and two doublets with $S^z =\pm 1$ and $S^z = \pm 2$.
However, as we will argue below, once we move away from the dimer
limit, they decay into excitations corresponding to $|t^{0,\pm}\rangle$.
Thus, for the purpose of classification of elementary
excitations, it is sufficient to consider the (split) triplet
$|t^{0,\pm}\rangle$.

Given these states in single dimers, the states in
the entire system is given as follows: The ground state is the state
that all dimers form $|s \rangle$.  The lowest excited state is the
state that $|t^+ \rangle$ or $|t^- \rangle$ is excited in one of the
dimers, while the other dimers remain in $|s \rangle$.  Similarly in the
next lowest excited state, $|t^0 \rangle$ is excited in one of the
dimers and the other dimers form $|s \rangle$.  Here, we denote these
states $|0 \rangle$, $|\textrm{one}~ t^+ \rangle$, $|\textrm{one}~ t^-
\rangle$, and $|\textrm{one}~ t^0 \rangle$, respectively.
In fact, in a system which consists of $N$ dimers ($2N$ sites),
there are $N$ degenerate states for each class of
$|\textrm{one}~ t^\zeta \rangle$, where $\zeta=0, \pm$,
corresponding to the location of the excited dimer.
Let us define $| \textrm{one}~ t^\zeta, j \rangle$, where
$\zeta = 0, \pm$, as the state with 
the $j$-th dimer (of sites $2j-1$ and $2j$)
as the only excited dimer.
We can also define their Fourier transform
\begin{equation}
 | \textrm{one}~ t^\zeta, \tilde{q} \rangle
 \equiv \frac{1}{\sqrt{N}} \sum_j e^{i 2 j \tilde{q} } 
|\textrm{one}~ t^\zeta, j \rangle,
\label{eq.onet_q}
\end{equation}
where $\zeta=0, \pm$, labeled by the crystal momentum $\tilde{q}$
defined in the reduced Brillouin zone~\eqref{eq.reducedBZ}.

In this paper, we are primarily interested in
the DSF around the antiferromagnetic wavevector, $q \sim \pi$.
Here, the matrix
elements of spin operators among these states are found to be:
\begin{align}
\langle \textrm{one}~ t^+, \tilde{q} |S^+_q|0\rangle &\neq 0, \nonumber
\\
\langle \textrm{one}~ t^-, \tilde{q} |S^-_q|0\rangle &\neq 0,
\\
\langle \textrm{one}~ t^0, \tilde{q} |S^z_q|0\rangle &\neq 0 \, , \nonumber
\end{align}
generically if $\tilde{q} \equiv q \mod{\pi}$, and
\begin{align}
& \text{the other matrix elements of $S^\mu_q$ } \nonumber \\
& \text{between $|0 \rangle$ and $|\textrm{one}~ t^\pm
\rangle$, or $|\textrm{one}~ t^0 \rangle$}
= 0. \nonumber
\end{align}

Now let us consider increasing $\alpha$ from zero.  With a
nonzero $\alpha$, an excited state in a single dimer can ``hop'' to the
neighboring sites, giving rise to a dispersion.
It should be noted that, in the dimerized limit,
the momentum basis~\eqref{eq.onet_q} as well as
the localized basis $|\textrm{one}~ t^\zeta, j \rangle$,
represents $N$ degenerate excited states.
However, when the interdimer interaction is introduced
with $\alpha >0$,
momentum basis~\eqref{eq.onet_q}, but not the localized basis 
gives a set of approximate eigenstates if $\alpha$ is sufficiently small.
Now the energy of the excited state
$|\textrm{one} t^\zeta, \tilde{q} \rangle$ depends on
the momentum $\tilde{q}$; the dependence is nothing but the
dispersion relation.
Since the structure of the triplet split into the doublet
and the singlet by the anisotropy is very similar between
the single dimer and the SG field
theory,
it would be natural to assume that the structure of the
excitation does not qualitatively change in $0 \le \alpha < \alpha_c$.
Under this assumption, the states in the dimer limit $\alpha=0$,
$|0\rangle$, $|\textrm{one}~ t^+ \rangle$, $|\textrm{one}~ t^- \rangle$
and $|\textrm{one}~ t^0 \rangle$ are considered to be smoothly
connected to, at $\alpha \sim \alpha_c$,
the ground state $|\textrm{GS} \rangle$, and the single
soliton state $|S, \tilde{q} \rangle$,
the single antisoliton state $|\bar{S}, \tilde{q} \rangle$,
the single breather state $|B_1, \tilde{q} \rangle$,
each with the momentum $\tilde{q}$.
If this is the case, the following selection rules should hold:
\begin{align}
\langle S, \tilde{q}|S^+_q|\textrm{GS}\rangle &\neq 0, \notag \\
\langle \bar{S}, \tilde{q} |S^-_q|\textrm{GS}\rangle &\neq 0,
\label{eq:nonzero_element_ntenp} \\
\langle B_1, \tilde{q} |S^z_q|\textrm{GS}\rangle &\neq 0 \, , \notag
\end{align}
generically if $\tilde{q} \equiv q \mod{\pi}$, and
\begin{align}
&\text{the other matrix elements of $S^\mu_q$} \nonumber \\
&\text{between $|\textrm{GS}\rangle$ and
$|S, \tilde{q} \rangle$,
$|\bar{S}, \tilde{q} \rangle$, or $|B_1, \tilde{q} \rangle$} 
=0
\nonumber
\end{align}
Assuming this, $S^+_q$, $S^-_q$ and $S^z_q$ respectively act
as the creation operators of the soliton, the antisoliton and the
breather.  Thus the soliton and the antisoliton contribute to the $x$,
$y$ components of the DSF and the breather does to the $z$ component.

\begin{figure}[tb]
\centering \includegraphics[width=3.25in,clip]{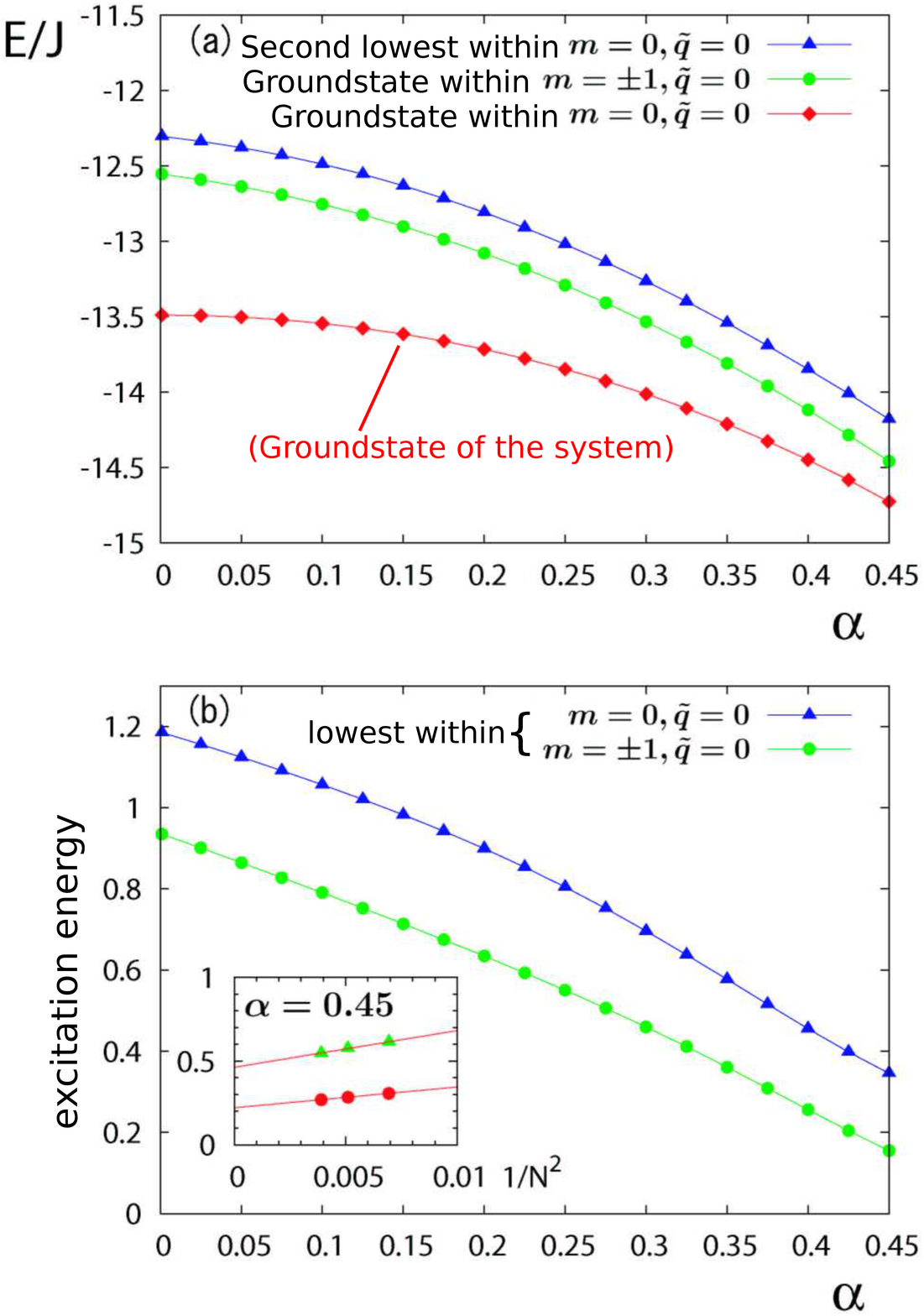} \caption{(a) The
$\alpha$-dependence of the energy of the ground state and the lowest
excited state in subspaces that the total magnetization $m=\sum_j S_j^z$
is $\pm1,0$ and the momentum $\tilde{q}=0$. ($N=16$)~ The case that
$\alpha=0.45$ corresponds to NTENP.  (b) The excitation energies
extrapolated to $N \to \infty$.  Inset: Extrapolation of the excitation
energies at $\alpha=0.45$ to $N \to \infty$.}
\label{fig:ene_enegap}\end{figure}

\begin{figure}[tb]
\centering \includegraphics[width=3.25in,clip]{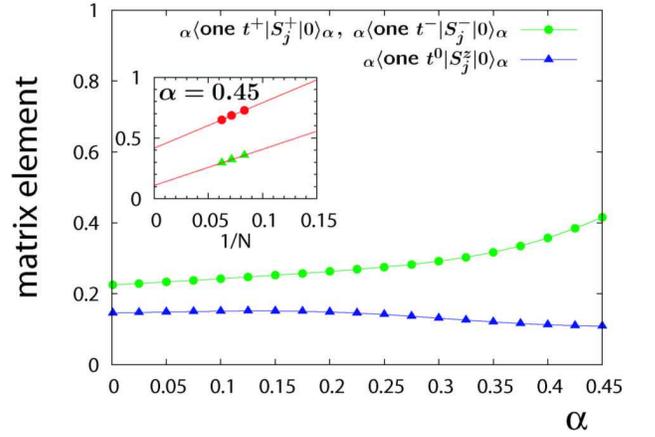} \caption{The
$\alpha$-dependence of the absolute value of the matrix elements
$_\alpha\langle \textrm{one}~ t^+, \tilde{q}=0|S^+_j|0\rangle_\alpha$,
$_\alpha\langle \textrm{one}~ t^-, \tilde{q}=0|S^-_j|0\rangle_\alpha$
and $_\alpha\langle \textrm{one}~ t^0,
\tilde{q}=0|S^z_j|0\rangle_\alpha$. ($N=16$)~ They correspond to
$\langle S|S^+_j|\textrm{GS}\rangle$, $\langle
\bar{S}|S^-_j|\textrm{GS}\rangle$ and $\langle
B_1|S^z_j|\textrm{GS}\rangle$ in the vicinity of the critical point
$\alpha=\alpha_c$, respectively.  Inset: Extrapolation of the matrix
elements at $\alpha=0.45$ to $N \to \infty$.}
\label{fig:matelement}\end{figure}

The assumption above is natural because we are interested in
a system in the dimer phase, and there is no phase transition between
the dimer limit $\alpha = 0$.  However, properties of excited states are
not necessarily the same, even if they are in the same phase and the
groundstates are adiabatically connected.  We confirmed the validity of
our assumption and that the matrix elements in
Eq.~\eqref{eq:nonzero_element_ntenp} does not vanish,
by the numerical exact
diagonalization calculation using the Lanczos method.  We show our
numerical results in Fig. \ref{fig:ene_enegap} and
Fig. \ref{fig:matelement}.
In what follows, we direct our attention to
the states with $\tilde{q}=0$ which correspond to the bottom of the each
branch in the energy spectrum, in order to compare with the preceding
results.~\cite{zheludev,hagiwara,suzukisuga}
As we will clarify later,
it suffices to obtain the
matrix element of the local spin operator $S^\alpha_j$
between the groundstate and the excited states.


Fig. \ref{fig:ene_enegap} (a) shows the energy of the ground state and
the lowest excited state in subspaces that the total magnetization
$m=\sum_j S_j^z$ is $\pm1,0$ and the momentum $\tilde{q}=0$.  This
figure is obtained by the exact diagonalization of the 16 site spin
system.  Fig. \ref{fig:ene_enegap} (b) shows the excitation energies
extrapolated from the calculated values in the 12, 14 and 16 site spin
systems to the infinite size.
These results show that the energy of the
elementary excitations changes adiabatically under the variation of
$\alpha$ from $0$ to $0.45$ ($\sim \alpha_c$), which is the value of
NTENP.  Thus our assumption that the excitation structure evolves
smoothly under the variation of $\alpha$, is confirmed.
In what follows, we represent the states at $\alpha$ that
are adiabatically connected
from $| \phi \rangle$ at $\alpha =0$, as $|\phi \rangle_\alpha$.  The
states $| \textrm{one}~ t^+ \rangle_{\alpha \sim \alpha_c}$, $|
\textrm{one}~ t^- \rangle_{\alpha \sim \alpha_c}$ and $| \textrm{one}~
t^z \rangle_{\alpha \sim \alpha_c}$ correspond to the soliton $|S
\rangle$, the antisoliton $|\bar{S} \rangle$ and the breather
$|B_1 \rangle$, respectively.  When $\alpha$ is finite, the elementary
excitations have a dispersion. Therefore we label them also with the
momentum $\tilde{q}$.

Fig. \ref{fig:matelement} shows the absolute value of the matrix
elements $_\alpha\langle \textrm{one}~ t^+,
\tilde{q}=0|S^+_j|0\rangle_\alpha$, $_\alpha\langle \textrm{one}~ t^-,
\tilde{q}=0|S^-_j|0\rangle_\alpha$ and $_\alpha\langle \textrm{one}~
t^0, \tilde{q}=0|S^z_j|0\rangle_\alpha$ in the variation of $\alpha$.
This figure is also obtained by the 16 site spin system exact
diagonalization.  From these calculated values, the matrix elements of
$S^\mu_q$ corresponding to them are evaluated as follows: As we will
discuss later in Sec. \ref{sec:raman}, the ground state $|0
\rangle$ is link-parity even,
and the states $|\textrm{one}~ t^+,
\tilde{q}=0 \rangle_\alpha$, $|\textrm{one}~ t^-, \tilde{q}=0
\rangle_\alpha$ and $|\textrm{one}~ t^z, \tilde{q}=0 \rangle_\alpha$ are
link-parity odd:
\begin{align}
P_l |0\rangle_\alpha &= |0\rangle_\alpha , \notag \\
P_l |\textrm{one}~ t^\zeta, \tilde{q}=0 \rangle_\alpha &=
-|\textrm{one}~ t^\zeta, \tilde{q}=0 \rangle_\alpha , \notag
\end{align}
where $\zeta=0,\pm$, and
the link parity operator $P_l$ represents a reflection with
respect to a link.  It satisfies $P_l^2 = I$, where $I$ is the identity
operator.  From these properties, it follows that
\begin{align}
_\alpha\langle \textrm{one}~ t^\zeta, \tilde{q}=0|S^+_{j+1}|0\rangle_\alpha
&=_\alpha\langle \textrm{one}~ t^\zeta, \tilde{q}=0|P_l S^+_j
P_l|0\rangle_\alpha \notag \\
&=- _\alpha\langle \textrm{one}~ t^\zeta, \tilde{q}=0|S^+_j|0\rangle_\alpha,
\label{eq:different_site}
\end{align}
where $\zeta=0,\pm$.
Accordingly,
$_\alpha\langle \textrm{one}~ t^\zeta, \tilde{q}=0|S^+_{q=\pi}|0\rangle_\alpha$
does not vanish, while 
$_\alpha\langle \textrm{one}~ t^\zeta, \tilde{q}=0|S^+_{q=0}
|0\rangle_\alpha =0$.
The adiabatic continuity of the excited states then implies
the same selection rule for the single soliton state
$|S, \tilde{q}=0 \rangle$, single antisoliton state
$|\bar{S}, \tilde{q}=0\rangle$, and 
single breather state $| B_1, \tilde{q}=0 \rangle$.
Thus we have confirmed the validity of our ansatz.

\subsection{Excitation structure of NTENP based on the SG field theory}

\begin{figure}[tb]
\centering \includegraphics[width=3in,clip]{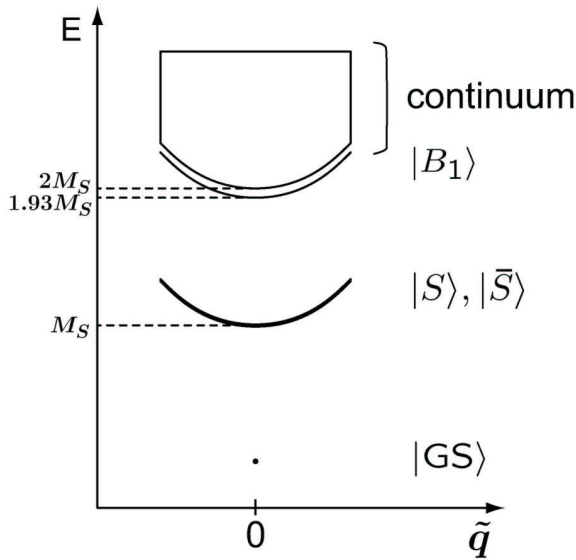} \caption{Schematic
view of the excitation spectrum of NTENP based on the SG field theory.
The doubly degenerated isolated mode correspond to the soliton and the
antisoliton is present and its lowest creation energy is $M_S$.  The
system has only the first breather and its lowest creation energy is
$1.93M_S$.  The excitation continuum starts from the $2M_S$ at the
momentum $\tilde{q}=0$.}  \label{fig:NTENPspectrum}
\end{figure}

Combining the results in Secs.~\ref{subsec:massratio}
and~\ref{subsec:dsf}, we obtain the following picture:
NTENP has three elementary excitations
that correspond to the soliton, the antisoliton and the first breather
in the SG theory. For these elementary excitations, the soliton mass and
the antisoliton mass are degenerate, and the first breather mass is
$1.93$ times the soliton mass.  (The mass degeneracy of
soliton/antisoliton is a common nature of
the SG field theories, that is, irrespective of the value of radius $R$.
On the other hand, the breather mass ratio depends on the value of the
radius $R$. For NTENP, $1/(2\pi R^2) \sim 1.7$.)
Thus, it is concluded that the doubly degenerated isolated mode is at
$M_S$ and the single isolated mode is at $1.93 M_S$ in the energy
spectrum of NTENP.  The excitation continuum starts from $2M_S$.  This
is schematically described in Fig. \ref{fig:NTENPspectrum}.

The soliton, the antisoliton and the breather are respectively created
by operating $S^+_q$, $S^-_q$ and $S^z_q$ on the ground state.
Conversely, when each spin operator creates a single elementary
excitation, its species is uniquely determined as soliton, antisoliton,
and breather.
Hence the single-particle peak due to the soliton/antisoliton
contribute only to
$\mathcal{S}^{xx}$ and $\mathcal{S}^{yy}$, and that due to
the breather does to only
$\mathcal{S}^{zz}$.  The intensity ratio between the lowest excited
states in $\mathcal{S}^{xx}$ and $\mathcal{S}^{zz}$ are evaluated from
the result in Fig. \ref{fig:matelement}.  Since the bottom of each
elementary excitation branch in the energy spectrum $\tilde{q}=0$
corresponds to $q=0$ and $q=\pi$, the intensity from the lowest excited
state in $\mathcal{S}^{\mu\mu}$ is the sum of
$\mathcal{S}^{\mu\mu}(0,\omega)$ and $\mathcal{S}^{\mu\mu}(\pi, \omega)$
(where $\omega$ is the energy transfer of the lowest excited state).
However, the matrix elements of $S^\mu_{q=0}$ vanishes due to the
relation~\eqref{eq:different_site} for each elementary excitations, then
only those of $S^\mu_{q=\pi}$ contribute.  For example, the contribution
of the soliton with $\tilde{q}=0$ occurs in
$\mathcal{S}^{xx}(\pi,\omega_S)$.  Taking these into consideration, the
intensity ratio are evaluated by
\begin{align}
\frac{|\langle B_1, \tilde{q}=0|S^z_{q=\pi} |\textrm{GS} \rangle|^2}
{\frac{1}{4}|\langle S, \tilde{q}=0|S^+_{q=\pi} |\textrm{GS} \rangle|^2
+ \frac{1}{4}|\langle \bar{S}, \tilde{q}=0|S^-_{q=\pi} |\textrm{GS}
\rangle |^2} \sim 0.22 .
\label{eq:intensity_ratio}
\end{align}
Here, we use the calculated values $_{0.45}\langle \textrm{one}~ t^+,
\tilde{q}=0| S^+_{q=\pi} | 0 \rangle_{0.45}$, $_{0.45}\langle
\textrm{one}~ t^-, \tilde{q}=0| S^-_{q=\pi} | 0 \rangle _{0.45}$ and
$_{0.45}\langle \textrm{one}~ t^z, \tilde{q}=0| S^z_{q=\pi}|0
\rangle_{0.45}$ in Fig. \ref{fig:matelement} for the soliton, the
antisoliton and the breather, respectively.


\section{Comparison with Preceding Results}

According to Refs.~\onlinecite{zheludev, hagiwara}, two main peaks
are present in the INS spectrum of NTENP.  The first peak occurs from
the fluctuation of spin $x$, $y$ components and the second from that of
$z$ component.  In an inelastic scan at the center of the Brillouin
zone, the energy ratio between the peak from spin $x$, $y$ component and
that from $z$ component is about $1.91/1.07 \sim 1.79$, and the
intensity of the second peak is about $0.2$ times that
of the first peak.
Furthermore, the DSF in the model~\eqref{eq:hamiltonian}
under a transverse magnetic field $H$
is numerically studied using the
continued-fraction method\cite{suzukisuga}.
Their result at $H=0$ can be compared with
the present study. The ratio between the energies
correspond to the first peak in
$\mathcal{S}^{xx}(\pi,\omega)$ and $\mathcal{S}^{zz}(\pi,\omega)$ is
$\sim 1.91$ and the intensity of the first peak of
$\mathcal{S}^{zz}(\pi,\omega)$ is about $0.22$ times that of
$\mathcal{S}^{xx}(\pi,\omega)$.
We also note that,
the ratio of the lowest gap
to $m=\pm 1$ and $m=0$ excitations
in our exact diagonalization
study, shown in Fig.~\ref{fig:ene_enegap},
is consistent with the predicted mass ratio $1.93$.
This gives an additional support
to our field-theory analysis.

The lowest-energy peak in the INS spectrum in
Refs.~\onlinecite{zheludev,hagiwara},
which corresponds\cite{suzukisuga} to
the lowest excited state in
$\mathcal{S}^{xx}(\pi,\omega)$,
is the contribution of the soliton/antisoliton in our picture.
Likewise, the second peak in the INS spectrum,
which corresponds to the lowest excited state in
$\mathcal{S}^{zz}(\pi,\omega)$,
is the contribution of the breather. 
Our results that the mass ratio between soliton and breather $M_B/M_S
\sim 1.93$ and that the creation operator of soliton, antisoliton and
breather is respectively $S^+_q$, $S^-_q$ and $S^z_q$, agree very
well with these results.
Our result~\eqref{eq:intensity_ratio} on the intensity ratio
also shows a good agreement.
Not only the peak positions but also the
polarization of the DSF is consistent.
Our field-theory approach thus provides a coherent theoretical
picture describing the preceding experimental and numerical results,
at least at zero magnetic field.
A slight disagreement with experiments may be attributed to the
following factors. 
First, we neglected the in-plane anisotropy $E$,
which must be present in NTENP with a triclinic crystal
structure.  Furthermore, the low-energy asymptotic 
description based on the SG field theory
is not exact for NTENP due to its finite energy gap.

More recently, ESR study of NTENP was reported in
Ref.~\onlinecite{Glazkov2010}.
We will make a brief comment on ESR in Sec.~\ref{sec:magfield}.

\section{Raman Scattering in NTENP}
\label{sec:raman}

As a new application of the present approach, in this section, 
we discuss the Raman scattering (RS) spectrum of NTENP and
predict its peak positions.  
Following our discussion in Sec.~\ref{subsec:dsf},
we first identify the non-vanishing matrix
elements in the dimer limit $\alpha =0$, and then extend the results
to $\alpha \neq 0$ using the continuity of the excitation
structure from $\alpha =0$ to $\alpha \lesssim \alpha_c$.

According to the theory of exchange-scattering in magnetic compounds,
the spectral function $I(\omega)$ for the scattered light at zero
temperature is given by ~\cite{parkinson}
\begin{equation}
I (\omega)=\sum_{n}\delta(\omega-(E_n-E_0))|\langle
\psi_n|\mathcal{H}_R|\psi_0 \rangle|^2 , \label{eq:selection rule
raman}\end{equation} where $| \psi_0 \rangle$ and $|\psi _n\rangle$ are
the ground state and the excited state of the Hamiltonian
(\ref{eq:hamiltonian}), whose energy eigenvalues are $E_0$ and $E_n$,
and $\mathcal{H}_R$ is the effective interaction for the
exchange-scattering of light,
\begin{equation}
\mathcal{H}_R=\sum_i(\bm{e}_i \cdot \bm{d}_{i ,i+1})(\bm{e}_s \cdot
\bm{d}_{i ,i+1})(\bm{S}_i \cdot \bm{S}_{i+1}) ,
\label{eq:raman}\end{equation} for the present one dimensional system
(\ref{eq:hamiltonian}).
Here, $\bm{e}_i$, $\bm{e}_s$ are unit
polarization vectors of incident and scattered light, and $\bm{d}_{i
,i+1}$ is a unit vector connecting the spin sites $i$ and $i+1$.  In
Eq. (\ref{eq:raman}), $\bm{d}_{i ,i+1}$ is constant in the present
system (\ref{eq:hamiltonian}); once the experimental apparatus is set
up, $\bm{e}_i$ and $\bm{e}_s$ are fixed.
(For a recent discussion of related problems, see
Ref.~\onlinecite{Sato-Raman}.)
Thus, we just need to know
matrix elements of the operator $\sum_i \bm{S}_i \cdot \bm{S}_{i+1}$ in
order to calculate RS spectrum.  This effective Raman
Hamiltonian~(\ref{eq:raman}), taking the form of an inner products of
spin operators, possesses several symmetries. Consequently, there are
following conserved quantities:
the total magnetization $m=\sum_j S_j^z$, the momentum and
the link parity of the states.  Link parity transformation is a
reflection with respect to a link.  In the systems with bond alternation
such as NTENP, the link parity is conserved but the site parity
(reflection with respect to a site) is not conserved.  The link parity
operator $P_l$ satisfies $P_l^2 = I$, where $I$ is the identity
operator.
As mentioned above, the selection rule for RS is quite different from
that for INS represented in Eq. (\ref{eq:dsf}).  This difference leads
to distinct peak positions between the INS and the RS spectrum of
NTENP.

In what follows, we discuss the overlaps between $|\psi_n \rangle$ and
$|\psi_0\rangle$ to detect the excited states that contribute to the RS
spectrum.  For this purpose, here we collect the conserved quantities in
the present system (\ref{eq:hamiltonian}) again: the total magnetization
$m=\sum_j S_j^z$, the momentum $\tilde{q}$ associated with the
invariance under the translation by two sites, and the link parity.

First we analyze the $\alpha = 0$ case.  The ground state $|0 \rangle$
is the eigenstate of $m$ with the eigenvalue $0$, while the lowest
excited states $|\textrm{one}~ t^+ \rangle$ and $|\textrm{one}~ t^-
\rangle$ are that with the eigenvalue $1$ and $-1$, respectively.  Since
$\mathcal{H}_R$ conserves the total magnetization $m$, these excited
states do not have overlap with $\mathcal{H}_R |0 \rangle$, and
do not contribute to the RS spectrum.  Thus in what follows, we only discuss
the subspace with $m=0$ in order to detect the states having
non-vanishing overlap with $\mathcal{H}_R |0 \rangle$.  The three
low-energy excited states in this subspace, applied the same notation as
the preceding one, are $|\textrm{one}~ t^0 \rangle$, $|\textrm{one}~
t^+, \textrm{one}~ t^- \rangle$, and $|\textrm{two}~ t^0 \rangle$, from
the lowest in energy, in the dimer limit $\alpha = 0$.  They correspond
to the single breather state, the scattering state of a soliton and an
antisoliton, and the scattering state of two breathers, respectively at
$\alpha \sim \alpha_c$.

The link parity of the ground state $|0 \rangle$ and these states are
calculated as follows: For a given state $|\phi \rangle$, we denote its
mapping by the link parity as $| \bar{\phi}\rangle = P_l|\phi \rangle$.
As a preparation, we first discuss the link parity of the states in a
single dimer.  The state $|s \rangle$ is link-parity even:
\begin{align}
|\bar{s} \rangle=P_l | s\rangle &= C_s P_l \, \bigl(|\uparrow \downarrow
\rangle+|\downarrow \uparrow \rangle+ \frac{4}{A-\sqrt{A^2+8}}
|00\rangle \bigr) \nonumber \\[-2pt] &= C_s (|\downarrow \uparrow
\rangle+|\uparrow \downarrow \rangle+ \frac{4}{A-\sqrt{A^2+8}}
|00\rangle ) \nonumber \\[-2pt] &= | s\rangle ,
\end{align}
where $A=2D-1$ and $C_s$ is the normalization constant.  In a similar
calculation, we find that the states $|t^0\rangle$, $|t^+\rangle$, and
$|t^-\rangle$ are link-parity odd:
\begin{align}
|\bar{t^0} \rangle=P_l | t^0\rangle &= -| t^0\rangle , \\ 
| \bar{t^+} \rangle= P_l | t^+\rangle &= -| t^+\rangle , \\
| \bar{t^-} \rangle= P_l | t^-\rangle &= -| t^-\rangle .
\end{align}

Based on these, we next discuss the states in the entire system.  We
note that $P_l$ changes the position of the dimers in addition to the
action on each dimer described above.  The link parity of $|0\rangle$ is
even:
\begin{align}
P_l | 0\rangle &= P_l\, \bigl(|s\rangle \cdots |s\rangle \bigr)
\nonumber \\[-1pt] &= |\bar{s}\rangle \cdots |\bar{s}\rangle \nonumber
\\[-1pt] &= |s\rangle \cdots |s\rangle = |0 \rangle .
\end{align}
Next we discuss the link parity of excited states in the subspace with
$m=0$.  A state with a fixed position of the excited dimer is not an
eigenstate of $P_l$ nor of the momentum.  For example,
the state $|\textrm{one}~ t^0 \rangle$
with the first dimer excited is transformed under $P_l$ as
\begin{equation}
P_l \, \bigl( |t^0\rangle |s\rangle |s\rangle \cdots |s\rangle \bigr) =
- |s\rangle |s\rangle \cdots |s\rangle |t^0\rangle . \nonumber
\end{equation}
On the other hand, momentum
eigenstates can be constructed as a linear combination of
\begin{align}
&|t^0\rangle |s\rangle |s\rangle \cdots |s\rangle , \hspace{3mm}
|s\rangle |t^0\rangle |s\rangle \cdots |s\rangle , \nonumber \\
&|s\rangle |s\rangle |t^0\rangle \cdots |s\rangle, \hspace{3mm} \cdots ,
\hspace{3mm} |s\rangle |s\rangle \cdots |s\rangle |t^0\rangle.
\nonumber
\end{align}
Since the ground state $|0 \rangle$ has zero momentum and
$\mathcal{H}_R$ is translationally invariant, only the zero-momentum
excited states contribute to the RS spectrum.  We thus only have to
examine the zero-momentum combination
\begin{align}
|\textrm{one}~ t^0, \tilde{q} =0 \rangle \hspace{3mm} \propto
 \hspace{3.2mm}&|t^0\rangle |s\rangle |s\rangle \cdots |s\rangle
 \nonumber \\ + \,&|s\rangle |t^0\rangle |s\rangle \cdots |s\rangle +
 \cdots \nonumber \\ + \,&|s\rangle |s\rangle \cdots |s\rangle
 |t^0\rangle.  \nonumber
\end{align}
It turns out that this is link-parity odd:
\begin{equation}
P_l |\textrm{one}~ t^0, \tilde{q} =0 \rangle = - |\textrm{one}~ t^0,
\tilde{q} =0 \rangle .
\end{equation}
Since the ground state $|0\rangle$ is link-parity even, $|\textrm{one}~
t^0, \tilde{q}=0 \rangle$ appears not to have the overlap with
$\mathcal{H}_R |0\rangle$.  Similarly, $|\textrm{one}~ t^+, \tilde{q}=0
\rangle$ and $|\textrm{one}~ t^-, \tilde{q}=0 \rangle$ are link-parity
odd, and do not have the overlap.

In the above analysis, 
we considered the dimerized limit $\alpha \rightarrow 0$.
However, the Hamiltonian respects the link parity
$P_l$ for any $\alpha$.
Thus, the adiabatic continuity of the excited states,
which was demonstrated numerically in Sec.~\ref{sec.dimer},
implies that the corresponding states
belong to the same eigenvalue of $P_l$ for any $\alpha < \alpha_c$.
It follows that, \emph{none}
of the ``single elementary excitation'' states
$|\textrm{one}~ t^\zeta \rangle_\alpha$ ($\zeta=0,\pm$)
(corresponding to a single
soliton, a single antisoliton and a
single breather for $\alpha \sim \alpha_c$)
contribute to the RS spectrum for $\alpha<\alpha_c$.

\begin{figure}[tb]
\centering \includegraphics[width=3.25in,clip]{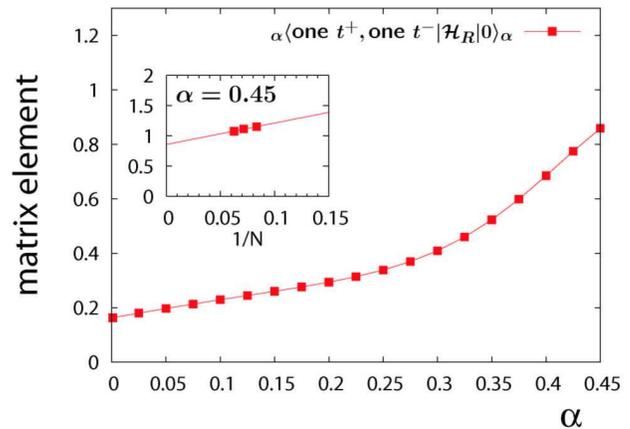} \caption{The
$\alpha$-dependence of the matrix element $_\alpha\langle \textrm{one}~
t^+, \textrm{one}~ t^- | \mathcal{H}_R |0\rangle_\alpha$. ($N=16$)~ The
state $| \textrm{one}~ t^+, \textrm{one}~ t^- \rangle_\alpha$
corresponds to the scattering state of a soliton and an antisoliton in
the vicinity of the critical point $\alpha=\alpha_c$.  Insets:
Extrapolation of the matrix element at $\alpha=0.45$ to $N \to \infty$.}
\label{fig:raman matelement}\end{figure}

Thus only the states in the excitation continuum have the possibility of
having non-vanishing overlap with $\mathcal{H}_R |0 \rangle$. 
In particular, the matrix element with the
``two elementary excitations'' state
$\langle \textrm{one}~ t^+, \textrm{one}~ t^- |\mathcal{H}_R |0 \rangle$
does not vanish in the dimer limit.
This transition is not forbidden by the selection rule due
to the link parity $P_l$.
Thus it is natural to expect that it remains non-zero for $\alpha < \alpha_c$.
Indeed, we examined the matrix element
$_\alpha\langle \textrm{one}~
t^+, \textrm{one}~ t^- | \mathcal{H}_R |0\rangle_\alpha$
numerically for $\alpha < \alpha_c$.
The result,
obtained after extrapolation to infinite size from
the systems with length $12, 14$ and $16$,
is shown in Fig.~\ref{fig:raman matelement}.
This confirms that the matrix element
does not vanish within the dimer phase, as expected.
The structure of excitations is continuous up
to the critical point $\alpha = \alpha_c$, also in this regard.

Thus the ``two elementary excitations'' state
$|\textrm{one}~ t^+, \textrm{one}~ t^- \rangle_\alpha$
should contribute to the RS spectrum of NENP.
In the vicinity of the critical point $\alpha \sim \alpha_c$,
this state may be regarded as the scattering state of a
soliton and an antisoliton.

\begin{figure}[tb]
\centering \includegraphics[width=2.5in,clip]{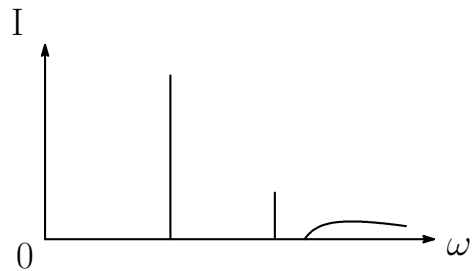}
\caption{Schematic view of the INS peak positions of NTENP. The first
peak corresponds to the soliton and the antisoliton, and the second peak
does to the breather.}  \label{fig:ins}\end{figure}

\begin{figure}[tb]
\centering \includegraphics[width=2.5in,clip]{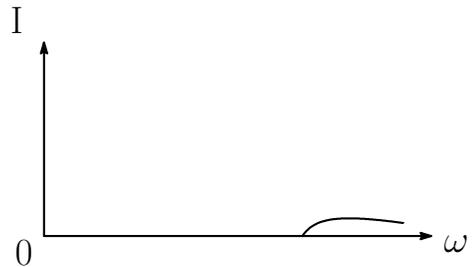}
\caption{Schematic view of the RS peak positions of NTENP. The sharp
peaks correspond to a single elementary excitation are absent due to the
selection rule.}  \label{fig:raman}\end{figure}

From these results, we predict that the sharp peaks corresponding to
creation of single elementary excitation
are absent in the RS spectrum of NTENP.
The RS spectrum consists only of excitation continuum, and
is qualitatively different from its INS spectrum.
This difference is schematically described in Fig.~\ref{fig:ins} and
Fig.~\ref{fig:raman}.

\section{Weak magnetic field}
\label{sec:magfield}

Finally, let us briefly discuss the effect of weak magnetic field.
First we consider magnetic field $H_z$
applied in $z$ direction.
Total magnetization $\sum_j S^z_j$ is then a conserved quantity.
Thus, the dispersion of each elementary excitation are
simply shifted according to its magnetization as
\begin{align}
\epsilon_S &= \sqrt{k^2 {v_s}^2 + {M_S}^2 } - H_z  & \mbox{(soliton)}, \\
\epsilon_{\bar{S}} &= \sqrt{k^2 {v_s}^2 + {M_S}^2 } + H_z  &
\mbox{(antisoliton)}, \\
\epsilon_{1} &= \sqrt{k^2 {v_s}^2 + {M_1}^2 } &
\mbox{(1st breather)} .
\end{align}
Phase transition occurs when the bottom of the soliton
dispersion touches zero at $H_z = M_S$.

Next we consider the effect of magnetic field $H_x$
applied perpendicular to the anisotropy axis $z$.
In this case, the total magnetization $\sum_j S^z_j$ is
no longer conserved.
Therefore, a rearrangement of elementary excitations would
occur.
The mixing may be described by the effective $3 \times 3$
Hamiltonian in the one-soliton, one-breather, one-antisoliton
subspace:
\begin{align}
\mathcal{H}_{\mbox{eff}} &= 
\left(
\begin{array}{ccc}
M_S & 0 & 0 \\
0 & b M_S & 0 \\
0 & 0 & M_S
\end{array}
\right)
-
\frac{H_x}{2}
\left(
\begin{array}{ccc}
0 & \sqrt{2} & 0 \\
\sqrt{2} & 0 & \sqrt{2} \\
0 & \sqrt{2} & 0
\end{array}
\right) ,
\end{align}
where $b$ is the mass ratio~\eqref{eq.massratio.NTENP}.

Diagonalizing this, we obtain the masses of 3
elementary excitations in the transverse magnetic field as
\begin{align}
M_A &= M_S, \\
M_\pm &= \frac{(1+b)M_S \pm \sqrt{{M_S}^2(b-1)^2+4 {H_x}^2}}{2} .
\end{align}
Namely, one of the elementary excitations (which is an antisymmetric
linear combination of soliton and antisoliton)
has constant mass $M_S$.
The other two masses depend nonlinearly on $H_x$.
The lowest mass $M_-$ is reduced to zero at
a critical field, where a phase transition occurs.

This is essentially identical to the preceding analysis
of anisotropic Haldane chains in Ref.~\onlinecite{Golinelli-JPhysCM1993}
which was applied to NDMAP~\cite{ndmap} and NTENP~\cite{hagiwara,Glazkov2010},
where the in-plane anisotropy $E$ is also taken into account.
As pointed out in Refs.~\onlinecite{hagiwara,suzukisuga},
the scattering intensity corresponding
to the higher-energy elementary excitation
(with the $M_+$) is suppressed in a moderate transverse field,
where the elementary excitation is absorbed into
the excitation continuum.
Systematic analysis of the effects of
the transverse field in the SG theory framework
is left for future.
This will be necessary for discussion of ESR spectra\cite{Glazkov2010}
in the present approach.

\section{Summary and Discussion}

In this study, we found that the low-energy excitations of NTENP are well
described in the framework of the SG field theory.  The elementary
excitations of NTENP correspond to the soliton, the antisoliton and the
breather, which are respectively created by applying $S^+_q$, $S^-_q$
and $S^z_q$ on the ground state.  Their correspondence with the original
spin model is deduced from the dimer limit $\alpha=0$, where the
original model is exactly solvable, and the numerical exact
diagonalization calculation which shows that the excitation structure is
smoothly connected from the dimer limit $\alpha =0$ to $\alpha \sim
\alpha_c$.  This picture of excitations well explains the preceding INS
experiments on NTENP and numerical calculations of the DSF. 

Based on the established picture, we found that the sharp
peaks, each of which corresponds to a single elementary excitation,
vanish in the RS spectrum of NTENP.
This is qualitatively distinct from the INS spectrum.
A slight disagreement with the experiments may be caused by the
neglect of the $E$ term anisotropy in the model
Hamiltonian~\eqref{eq:hamiltonian}, and the use of the
effective field theory, which is only asymptotically exact
in the low-energy limit, 

Although our primary focus in this paper was NTENP,
the present result could be applied to other systems
in the dimer phase of the Hamiltonian~\eqref{eq:hamiltonian}.
Depending on the value of the uniaxial anisotropy $D$,
the number of breathers changes:  The system has two breathers
for $0 \le D \lesssim 0.1J$, one breather for
$0.1J \lesssim D \lesssim 0.5J$
and no breather for $0.5J \lesssim D$.

Besides the bond-alternating chain~\eqref{eq:hamiltonian},
there are also completely different class of systems
that are described by the SG field theory.
An example in quantum magnetism is the $S=1/2$ antiferromagnetic chain with
a staggered field.  It describes several materials in a magnetic field,
such as Cu benzoate.  While this system is physically quite different
from the model discussed in the present paper, they are both described
by SG theory.~\cite{cubenzoate} The difference appears in the
correspondence between the spin operators and the boson field.  In fact,
in the staggered field case, soliton and antisoliton couple to $S^z$
while $n$-th breather couples to $S^x$ if $n$ is even, and to $S^y$ if
$n$ is odd.  (Here we define the direction of the staggered field as $x$
direction.)  As a consequence, observable spectra are quite different
from the S=1 chain studied in the present paper.

As a final remark, in the context of the SG field theory, difference
between the dimer phase and the Haldane phase is nothing but the sign of
the cosine term in the Lagrangian density (\ref{eq:sine-Gordon}).
This
change should not yield a qualitative difference in the properties of the
SG theory, although its physical consequences may be altered. 
The SG field theory could provide a starting point
for a unified understanding of dynamics in different phases
of $S=1$ antiferromagnetic chains.
In a related but somewhat different context, anisotropic $S=1$
chains without bond alternation
was studied in terms of SG theory.\cite{CamposVenuti2006}
It would be interesting to compare the two systems
in more detail.

\section*{Acknowledgment}
The authors thank, in particular,
Shunsuke C. Furuya, Sei-ichiro Suga, and Takafumi Suzuki
for valuable discussions which were essential to complete
the present work.
The authors are also grateful to
Masayuki Hagiwara, 
Yuhei Natsume, Hidetoshi Nishimori, Kiyomi Okamoto,
and Shintaro Takayoshi
for useful comments.
The numerical calculations in this paper are based on KOBEPACK version 1.0
by Takashi Tonegawa, Makoto Kaburagi and Tomotoshi
Nishino.  This work is supported in part by a Grant-in-Aid for
scientific research (KAKENHI) Nos. 21540381 and 20102008,
and by the 21st Century COE program at Tokyo
Institute of Technology ``Nanometer-Scale Quantum Physics'', both from
MEXT of Japan.

\end{document}